\documentclass[prb,preprint,showpacs,superscriptaddress]{revtex4}
\usepackage{graphicx}
\begin{document}
\title{Neutron Irradiation of SmFeAsO$_\mathrm{1-x}$F$_\mathrm{x}$}

\author{M~Eisterer}
\author{H~W~Weber}

\address{Atominstitut der \"Osterreichischen Universit\"aten, Vienna University of Technology, 1020
Vienna, Austria}

\author{J~Jiang}
\author{J~D~Weiss}
\author{A~Yamamoto}
\author{A~A~Polyanskii}
\author{E~E~Hellstrom}
\author{D~C~Larbalestier}

\address{National High Magnetic Field Laboratory, Florida State University, Tallahassee,
FL 32310, USA }


\begin{abstract}
SmFeAsO$_\mathrm{1-x}$F$_\mathrm{x}$ was irradiated in a fission
reactor to a fast ($E>0.1$\,MeV) neutron fluence of $4\times
10^{21}$\,m$^{-2}$. The introduced defects increase the normal
state resistivity due to a reduction in the mean free path of the
charge carriers. This leads to an enhancement of the upper
critical field at low temperatures. The critical current density
within the grains, $J_\mathrm{c}$, increases upon irradiation. The
second maximum in the field dependence of $J_\mathrm{c}$
disappears and the critical current density becomes a
monotonically decreasing function of the applied magnetic field.

\end{abstract}


\maketitle

\newpage

\section{Introduction}

The discovery of superconductivity in the iron pnictides
\cite{Kam08} is not only interesting from a theoretical point of
view, but this new class of superconductors could also become
important for applications. The fundamental superconducting
parameters are promising. The transition temperature reaches about
55\,K \cite{Ren08,Ren08b,Yan08}, which is not as high as in most
cuprates, but significantly higher than in the technologically
relevant superconductors NbTi and Nb$_3$Sn or in MgB$_2$. The
upper critical field, $B_\mathrm{c2}$ is extremely large
($>50$\,T) \cite{Che08,Gao08,Jar08,Jia08,Sas08,Sen08b} and thermal
fluctuations seem to be less important than in the cuprates
\cite{Jar08,Yam09}, where loss free currents are restricted to
fields far below the upper critical field, at least at elevated
temperatures. Zero resistivity was demonstrated at fields close to
$B_\mathrm{c2}(T)$ in pnictide single crystals \cite{Jia08}.

Irradiation techniques are a powerful tool for assessing the
influence of defects on superconductors, because they allow
investigating the same sample prior to and after the irradiation,
which excludes problems of sample to sample variations. In
particular neutron irradiation was used in extended studies of the
influence of disorder in MgB$_2$
\cite{Tar06,Wil06,Kru07,Eis07rev}, including the demonstration of
the disappearance of two band superconductivity due to interband
scattering at high levels of disorder \cite{Put06}. An increase in
upper critical field and a reduction of its anisotropy were
reported and are expected theoretically. A similar behaviour of
$B_\mathrm{c2}$ was also found in the A15 compounds \cite{Put08}
and a reduction of anisotropy was reported for the cuprates
(Hg-1201) \cite{Zeh05}.

The neutron induced defects are also highly suitable for
investigating flux pinning. This was done successfully in V$_3$Si
\cite{Mei85}, the cuprates \cite{Sau98,Wer00} and MgB$_2$
\cite{Zeh04}.

The present contribution reports on a first neutron irradiation
experiment on the new FeAs based superconductors. Changes of the
reversible and irreversible superconducting properties were found
to be similar to other superconducting materials.

\section{Experimental\label{secexp}}

The SmFeAsO$_\mathrm{1-x}$F$_\mathrm{x}$ sample was prepared at
the National High Magnetic Field Laboratory. The starting
materials of As, Sm, Fe, Fe$_2$O$_3$ and SmF$_3$ were mixed and
pressed into a pellet, wrapped with Nb foil, and sealed in a
stainless steel tube. The sealed sample was heat treated at
1160\,$^\circ$C for 6 hours in a high temperature isostatic press
under a pressure of 280\,MPa. The main phase of the sample is
SmFeAsO$_\mathrm{1-x}$F$_\mathrm{x}$, with a grain size of 10 to
15 micrometers. The impurity phases include SmAs, SmOF and FeAs.
The size of the sample used for this study was $1.1\times 1.7
\times 3.7$\,mm$^3$.

Neutron irradiation was performed in the central irradiation
facility of the TRIGA-Mark-II reactor at the Atomic Institute in
Vienna. The sample was sealed into a quartz tube and exposed to
the neutron flux for 14 hours and 38 minutes, corresponding to a
fast ($E>0.1$\,MeV), thermal ($E<0.55$\,eV), and total neutron
fluence of $4\times 10^{21}$\,m$^{-2}$, $3.2\times
10^{21}$\,m$^{-2}$, and $1.1\times 10^{22}$\,m$^{-2}$,
respectively \cite{Web86}. Neutrons transfer their energy to the
lattice atoms by direct collisions. The transferred energy must
exceed the binding energy of the lattice atom to displace it, thus
only fast neutrons lead to defects. No indirect defect producing
mechanism (e.g. an induced $\alpha$-emission in MgB$_2$
\cite{Eis02,Tar06} or a fission reaction in uranium doped
YBa$_2$Cu$_3$O$_{7-\delta}$ \cite{Wei98}) exists.

The smallest resulting defects are single displaced atoms (point
defects), but also larger defects might occur. In high temperature
superconductors, the largest defects are so called collision
cascades \cite{Fri93}. These spherical defects are amorphous with
a diameter of 2 to 3\,nm, the surrounding strain field enlarges
the defect to about 5\,nm.  Similar defects were also found in
MgB$_2$ \cite{Zeh04,Mar08}. The actual size, morphology, and
density of the defects in SmFeAsO$_{1-x}$F$_x$ are currently
unknown. However, the defects should be randomly generated,
leading to a homogeneous defect density on a macroscopic length
scale. Self shielding effects can be neglected, since the
penetration depth of fast neutrons is estimated to be a few
centimeters, which is much larger than the sample dimensions. Only
neutrons of low or intermediate energies are shielded efficiently,
because of the large neutron cross section of samarium at these
energies, but the corresponding reactions or collisions are not
expected to produce any defects.

The resistivity was measured at various fixed fields while cooling
at a rate of 10\,K/h with an applied current of 10\,mA. Current
and voltage contacts were made by silver paste. The distance
between the two voltage contacts was about 1\,mm.

The transition temperature at each field, $T_\mathrm{c}(B)$, was
defined as the temperature, where the resistivity drops to
$0.95\rho_\mathrm{n}(T)$ ($\rho_\mathrm{n}(T)$ was extrapolated
linearly from its behaviour between 55 and 60\,K). The upper
critical field, $B_{c2}(T)$, was obtained by inversion of
$T_\mathrm{c}(B)$. The transition width, $\Delta T$, is defined by
the difference between $T_\mathrm{c}$ and the temperature where
$\rho(T)$ becomes $0.05\rho_\mathrm{n}(T)$.

Magnetization loops at various temperatures were recorded in a
commercial 7\,T SQUID magnetometer. The field was always oriented
parallel to the smallest sample dimension (transverse geometry).
The ac susceptibility at 33\,Hz was measured with an amplitude of
30\,$\mu$T at various fields and temperatures in order to estimate
the shielding fraction. The demagnetization factor was calculated
numerically for the actual sample geometry.

\section{Results and Discussion}

\begin{figure} \centering \includegraphics[clip,width=0.5\textwidth]{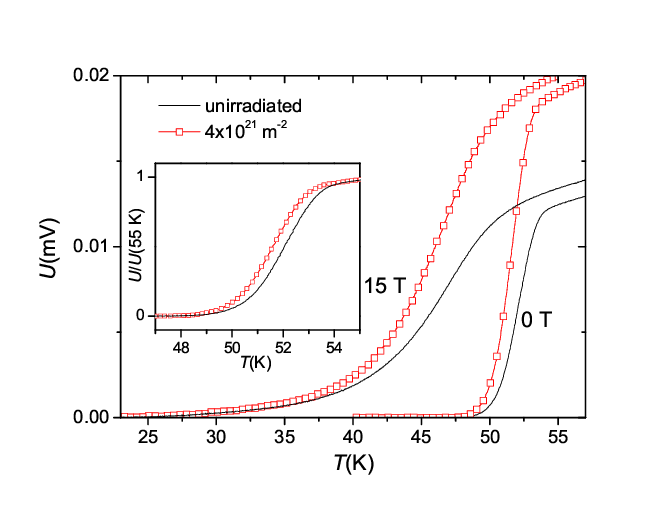}
\caption{Resistive transitions prior to and after neutron
irradiation. The insert shows the zero field transition normalized
to the voltage at 55\,K. The transition broadens with magnetic
field, but no broadening results from the irradiation. The
resistance is significantly enhanced by irradiation}
\label{FigResTran}
\end{figure}

The transition temperature decreases after irradiation, from
53.6\,K to 53.2\,K, which is similar to the decrease found in
YBa$_2$Cu$_3$O$_{7-\delta}$ \cite{Sau91} but less than in the
thallium-based cuprate superconductors \cite{Bra97} at the same
fluence. This decrease is ascribed to d-wave superconductivity in
the cuprates \cite{Mil88,Rad93}, the reason for the reduction in
$T_\mathrm{c}$ of SmFeAsO$_\mathrm{1-x}$F$_\mathrm{x}$ is
currently unknown. A moderate amount of non-magnetic impurities
does not reduce $T_\mathrm{c}$ in isotropic single band
superconductors, thus the anisotropy \cite{Rui06,Sch05,Zeh03} or
multiband superconductivity \cite{Gol97,Erw03,Dol05,Put07} are
possible candidates for explaining this decrease. Recently, a
suppression of $T_\mathrm{c}$ by impurity scattering was also
predicted for extended s-wave superconductors \cite{Ban09}.

The transition width is rather large ($\Delta T=3.8$\,K at 0\,T)
and significantly broadens in magnetic fields, as can be seen in
Fig.\ref{FigResTran}. It remains unchanged after irradiation,
which confirms the homogeneous distribution of the introduced
defects.

The measured resistivity increases due to the introduction of the
defects, from about 235\,$\mu\Omega$cm to 355\,$\mu\Omega$cm
($\sim$50\%) at 55\,K.  We also observe an increase of the phonon
contribution to the resistivity between 50\,K and 300\,K,
$\Delta\rho:=\rho_\mathrm{n}(300\,\mathrm{K})-\rho_\mathrm{n}(55\,\mathrm{K})$,
from 470\,$\mu\Omega$cm to 650\,$\mu\Omega$cm, which is expected
to remain unchanged in single band conductors, but can be altered
in multi-band conductors \cite{Maz02} (e.g.
SmFeAsO$_\mathrm{1-x}$F$_\mathrm{x}$). Alternatively, the increase
in $\Delta\rho$ could result from a reduced connectivity
\cite{Row03} or simply from changes in the distance between the
voltage contacts, which had to be removed for the irradiation and
renewed afterwards. A simple correction of these effects (keeping
$\Delta\rho$ constant) reduces the increase in resistivity due to
the irradiation to about 10\%. The experimental accuracy (distance
between voltage contacts) does not allow a final conclusion on the
absolute change of the resistivity or which scenario applies. A
reliable indication of the change in resistivity is given by
$\rho_\mathrm{n}(300\,\mathrm{K})/\rho_\mathrm{n}(55\,\mathrm{K})$,
which is independent of possible geometrical errors and which
decreases from 3.03 to 2.83. Thus, we conclude that the residual
resistivity is enhanced. Note that the resistivity is strongly
temperature dependent above $T_\mathrm{c}$. Therefore, the
residual resistivity $\rho_0$ is smaller than
$\rho_\mathrm{n}(55\,\mathrm{K})$ but cannot be assessed. The
residual resistivity ratio
($\mathrm{RRR}=\rho_\mathrm{n}(300\,\mathrm{K})/\rho_0$) is
certainly larger and might be changed much more significantly due
to the irradiation. This is also true for the relative change in
residual resistivity. We emphasize that the resistivity and the
RRR might be influenced by the impurity phases.

\begin{figure} \centering \includegraphics[clip,width=0.5\textwidth]{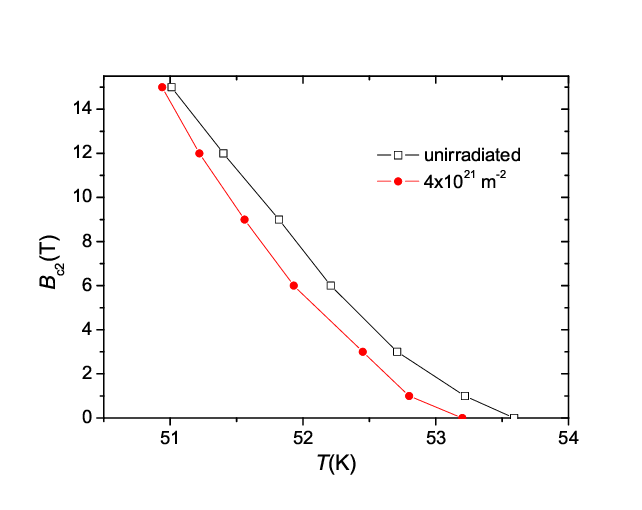}
\caption{Change of the upper critical field after neutron
irradiation} \label{FigBc2}
\end{figure}

A decrease in the mean free path of the charge carriers is
expected to result in an increase of the upper critical field.
Indeed, the slope $-\mathrm{d}B_\mathrm{c2}/\mathrm{d}T$ increases
after irradiation (Fig.~\ref{FigBc2}), but this cannot compensate
the small reduction ($\sim 0.4$\,K) of the transition temperature
in the investigated field range ($\leq15$\,T). However, the
enhanced slope indicates that the upper critical field will be
enhanced at low temperatures. The slope is nearly constant above
6\,T ($-\mathrm{d}B_\mathrm{c2}/\mathrm{d}T=7.5$\,T/K) before the
irradiation and a positive curvature can be observed up to 15\,T
after the irradiation, with a maximum slope of 10.7\,T/K between
12 and 15\,T. This corresponds to an enhancement by about 40\,\%.
Note that this enhancement is incompatible with d-wave
superconductivity, for which impurity scattering is predicted to
\emph{decrease} $-\mathrm{d}B_\mathrm{c2}/\mathrm{d}T$
\cite{Won94}.

\begin{figure} \centering \includegraphics[clip,width=0.5\textwidth]{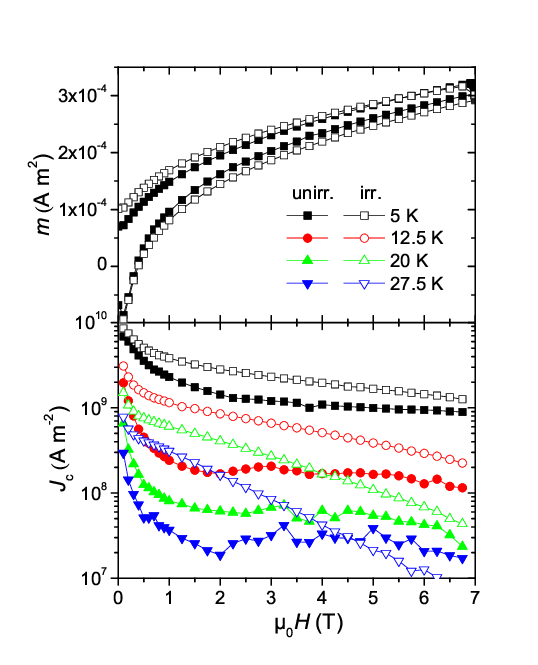}
\caption{The width of the hysteresis loop increases after
irradiation (upper panel) and the ``fishtail'' effect disappears
(lower panel).} \label{Figloopmirr}
\end{figure}

The hysteresis loop at 5\,K prior to and after neutron irradiation
is plotted in the upper panel of Fig.~\ref{Figloopmirr}. The
paramagnetic background, which arises most probably from impurity
phases, slightly decreases, but the hysteresis increases. The
latter indicates an increase of the critical current densities,
which are estimated from the Bean model in the lower panel. Zero
intergranular currents (see below) were assumed and cuboidal
grains with typical dimensions of 10\,$\mu$m, as found by
backscattered electron imaging. The solid and open symbols refer
to the unirradiated and irradiated state, respectively. The
critical current density, $J_\mathrm{c}$, increases more
significantly at higher temperatures and the ``fishtail'' effect
\cite{Che08,Yam08} disappears after neutron irradiation, i.e.
$J_\mathrm{c}$ decreases monotonically with field. This is
strikingly similar to the changes found in single crystalline
\cite{Wer00,Wis00} and melt textured \cite{Eis98,Gon02b} cuprate
superconductors and indicates that pinning in the unirradiated
sample is between weak (ordered flux line lattice) and strong
(disordered flux line lattice) pinning. Thus, an order-disorder
transition takes place at intermediate fields which leads to the
second peak (or fishtail effect) in the magnetization curve
\cite{Kha96,Gia97,Gam98,Mik01,Zeh04}. The neutron-induced defects
are strong enough to deform the flux line lattice plastically over
the whole field range. However, the defects resulting from neutron
irradiation are usually not the most efficient pinning centers.
For instance, the critical current densities in neutron-irradiated
cuprates are approximately one order of magnitude smaller than the
highest reported values (in films \cite{Fol07}, or in single
crystals containing columnar defects \cite{Civ97}). Thus, the
critical currents should be more efficiently improved in
SmFeAsO$_\mathrm{1-x}$F$_\mathrm{x}$ by the addition of stronger
pinning centres.

\begin{figure} \centering \includegraphics[clip,width=0.5\textwidth]{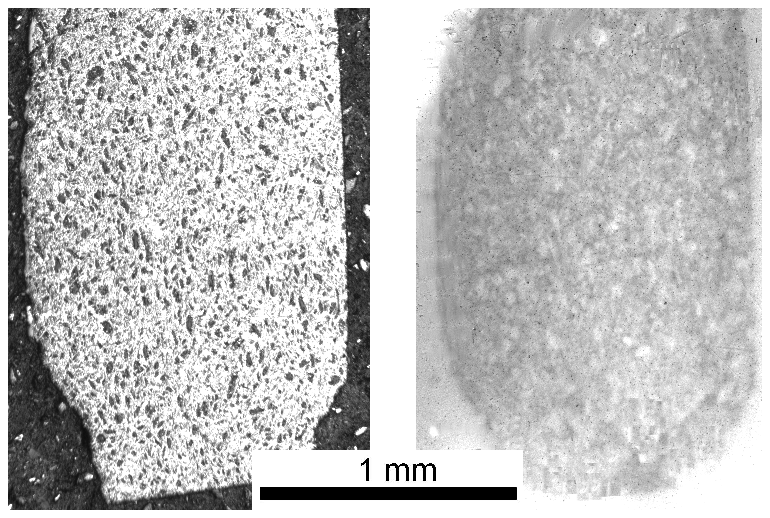}
\caption{Optical (left) and magneto-optical (right) image of a
reference sample at 6\,K and 60\,mT after zero field cooling. No
global Bean profile is established, showing that the intergranular
currents are negligible.} \label{FigMO}
\end{figure}

The interpretation of the irreversible magnetic moment,
$m_\mathrm{irr}$, in terms of $J_\mathrm{c}$ is quite delicate,
since it is a priori unknown, whether $m_\mathrm{irr}$ is
predominantly generated by (intergranular) currents flowing around
the whole sample, or by (intragranular) currents shielding
individual grains only. We observed a pronounced magnetic
granularity by magneto-optical imaging, with no global Bean
profile (Fig.~\ref{FigMO}, right). Thus, it is safe to assume that
the irreversible magnetic moment is made up nearly entirely by
intragranular currents. This behavior is in marked contrast to
that seen in a high pressure treated Sm-1111 sample where clear
evidence was seen of the global current flow \cite{Yam08,Kat09}.

\begin{figure} \centering \includegraphics[clip,width=0.5\textwidth]{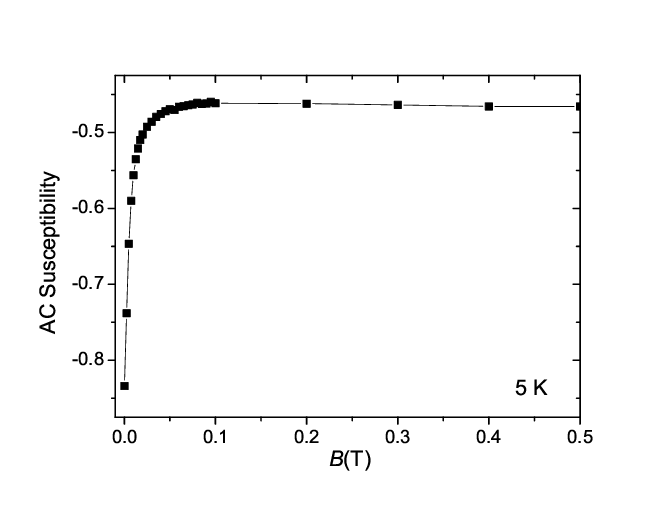}
\caption{The ac susceptibility decreases rapidly with increasing
magnetic field, since the grains decouple.} \label{FigAc}
\end{figure}

This is also demonstrated by ac susceptibility measurements.
Diamagnetic shielding was nearly perfect in the present sample at
zero magnetic field. The data in Fig.~\ref{FigAc} refer to 5\,K.
The susceptibility was calculated by assuming a demagnetization
factor of 0.58. The slightly smaller value at zero applied dc
field (-0.83 instead of -1 in the ideal case) can result either
from a residual magnetic field, or from geometrical imperfections.
The ac susceptibility decreases by a factor of nearly 2 by
applying a dc field of only 60\,mT and remains constant at higher
fields. The rapid decrease at low fields reflects the decoupling
of the individual grains and the small ac field penetrates the
whole sample along the grain boundaries at fields above about 60\,
mT, when the intergranular currents become negligible.

\section{Conclusions}

The introduction of disorder by fast neutron irradiation enhances
the normal state resistivity and the upper critical field at low
temperatures. Efficient flux pinning centers are introduced, which
enhance $J_\mathrm{c}$. Like in the cuprates, the second maximum
(``fishtail'') in the field dependence of the critical current
disappears, implying that the irradiation-induced defects distort
the flux line lattice plastically at all magnetic fields.

\begin{acknowledgments}

We are grateful to A. Gurevich, M. Putti, C. Tarantini and F.
Kametani for discussions. Work at the NHMFL was supported by NSF,
the State of Florida, and AFOSR.

\end{acknowledgments}



\end{document}